# A Pattern Recognition Method for Partial Discharge Detection on Insulated Overhead Conductors

Ming Dong
Department of System Planning and Asset Management
ENMAX Power Corporation
Calgary, Canada
mingdong@ieee.org

Jessie Sun
Auroki Analtyics
Calgary, Canada
sunzexi66@gmail.com

Carl Wang
Department of Power System Studies
Altalink L.P.
Calgary, Canada
carl.wang@altalink.ca

*Abstract*—Today, insulated overhead conductors are increasingly used in many places of the world due to the higher operational reliability, elimination of phase-to-phase contact, closer distances between phases and stronger protection for animals. However, the standard protection devices are often not able to detect the conductor's phase-to-ground fault and the more frequent tree/tree branch hitting conductor events as these events only lead to partial discharge (PD) activities instead of causing overcurrent seen on bare conductors. To solve this problem, in recent years, Technical University of Ostrava (VSB) devised a special meter to measure the voltage signal of the stray electrical field along the insulated overhead conductors, hoping to detect the above hazardous PD activities. In 2018, VSB published a large amount of waveform data recorded by their meter on Kaggle, the world's largest data science collaboration platform, looking for promising pattern recognition methods for this application. To tackle this challenge, we developed a unique method based on Seasonal and Trend decomposition using Loess (STL) and Support Vector Machine (SVM) to recognize PD activities on insulated overhead conductors. Different SVM kernels were tested and compared. Satisfactory classification rates on VSB dataset were achieved with the use of Gaussian radial basis kernel.

## I. INTRODUCTION

Insulated overhead conductors (IOC) or covered overhead conductors have been used in many places around the world. Today, some bare conductors are being replaced with insulated overhead conductors due to the following benefits from IOC [1]:

- Using IOC can avoid phase-to-phase fault and enhance system reliability. IOC is especially useful in areas with hazardous weather conditions such as strong wind where short circuits often happen between adjacent phases;
- Using IOC can reduce physical distances between phase conductors and physical distances between phase conductors and the towers or poles. For bare conductor systems, it is necessary to maintain minimum approach distance for safety and reliability reasons while for IOC, this requirement can be relaxed;
- Using IOC can reduce the fault occurrence caused by tree or tree branch hitting the conductors;
- Using IOC can reduce animal caused faults and protect animals.

However, IOC has two disadvantages related to protection: first, the standard protection devices used for bare conductor systems are often not able to detect the IOC's phase-to-ground fault. This is because with the insulation cover, the phase-to-ground fault will not likely cause an overcurrent when IOC breaks and falls to the ground. This can create risky situation to people in close proximity to the fallen conductors [2-3].

Similarly, the tree or tree branch hitting the conductors will not be detected by the upstream protection devices. This can become a potential reliability threat in some locations where the tree and tree branches frequently hit the conductors due to wind or continuously push and bend the conductors. Eventually, the power line will be damaged, causing a power outage or starting a tree fire [2-3].

Although being invisible to standard protection devices, the above two types of events can cause partial discharge (PD) activities [2-4]. Therefore, if the PD activities can be detected, these events can be detected and proper preventative actions can be taken.

In order to acquire PD signals, Technical University of Ostrava (VSB) in Czech Republic has made tremendous effort in the past few years [4-6]. They designed a measurement device that can be attached to the jacket of IOC to measure PDs. The setup is shown in Fig.1. A single layer coil is wrapped around the IOC to acquire the stray electrical field voltage along the IOC. The voltage signal is further taken by using a capacitance voltage divider whose output capacity, along with an inductor, is connected in parallel with the voltage output terminals [4]. Compared to another possible solution which is directly measuring the current in the conductor using Rogowski sensor [3], VSB's method is more cost-effective.

In addition to the development of the measurement device, VSB also worked on developing signal processing algorithms to detect PD activities caused by phase-to-ground fault and tree events [6]. In 2018, VSB published a dataset which contains a large number of waveform measurements recorded using the above device. The dataset was published on Kaggle, the world's largest data science collaboration platform [7]. The intent is to attract worldwide researchers to study the waveform data and develop functional pattern recognition algorithms for PD detection.

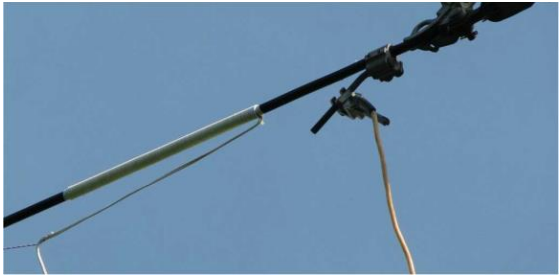

(a) A single layer coil wrapped around the IOC

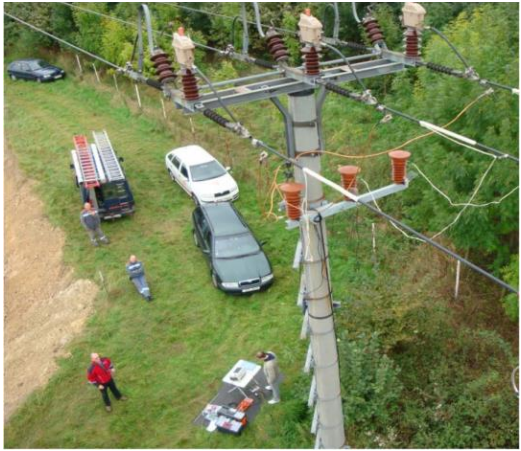

(b) Coils connected to three phases of a medium-voltage power line

Fig. 1. The setup of VSB's PD measurement device for insulated overhead conductors

To tackle this challenge, this paper proposes a method shown in Fig.2: first, Seasonal and Trend decomposition using Loess (STL) is used as a time-series analysis technique to decompose every raw voltage signal into three components, i.e. trend, seasonal and residual components. The proposed method then only focuses on the residual components of each signal where irregular noise information is most abundant. The method then produces a few data features from each residual component. This process is applied a few times with different seasonal window lengths. Eventually, the data features from different STLs are merged and normalized together. Data oversampling is also adopted to balance the records with PD activities and the records with no PD activities. In the end, the features and the known statuses are fed into a Support Vector Machine (SVM) classifier for training. The trained classifier will be able to detect future PDs.

This paper firstly introduces the VSB dataset. It then briefly explains the theories of STL and SVM. In the end, the proposed method is applied to the VSB dataset and detailed classification results for different SVM kernels are presented and discussed. The results show that the proposed pattern recognition method is promising for PD detection on IOC using VSB type of measurement devices.

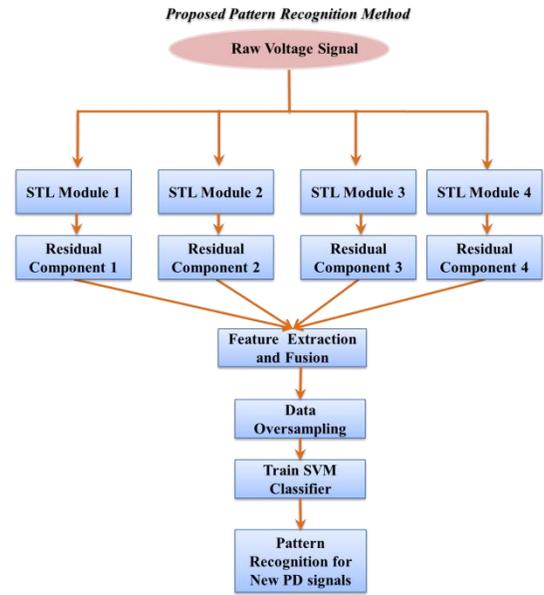

Fig. 2. The flowchart of the proposed method

## II. THE VSB DATASET

The VSB dataset contains 8,711 pre-labeled voltage signals recorded by their measurement device introduced in Section I. These signals are three-phase signals acquired in 2,903 times. Each signal has been pre-labeled as either having PD (525) or no PD (8,186). Each signal is a one-cycle voltage waveform (at 50Hz). It contains 800,000 data points in each waveform. Due to the large size of the data, Hadoop Distributed File System (HDFS) format is used. Special data manipulation is required to access the signals before processing. Sample waveforms of a typical PD record and non-PD record are plotted in Fig.3

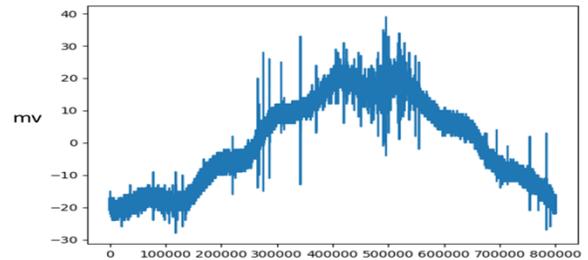

(a) non-PD signal

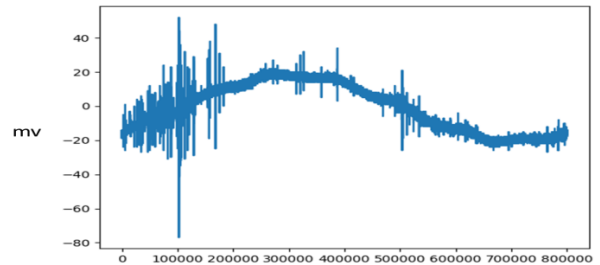

(b) PD signal

Fig. 3. Sample signal records from VSB dataset

## III. THE THEORIES OF STL AND SVM

This section will briefly review two key techniques used in the proposed pattern recognition method: STL and SVM.

### A. Seasonal and Trend decomposition using Loess

STL is a widely-used time series decomposition technique [8]. It intends to decompose a time-series into the three components given as below:

$$y_t = T_t + S_t + I_t \quad (1)$$

where $y_t$ is the target time series for decomposition. $T_t$, $S_t$ and $I_t$ are the trend component, seasonal component and residual component respectively. Their meanings are explained as below:

- $T_t$ reflects the long-term progression in the raw signal (secular variation). A trend exists when there is a persistent increasing or decreasing direction in the data.
- $S_t$ reflects seasonality (seasonal variation) in the raw signal. A seasonal pattern is the repeated cyclic pattern observed once in a while. $S_t$ can change over time.
- $I_t$ reflects irrgularity or noise in the raw signal. It is related to random and irregualr influences. It is also the residual part after excluding $T_t$ and $S_t$ from $y_t$. By nature, PD is a random and irregular component and this is the reason the proposed method focuses on the characteristics of residual component only.

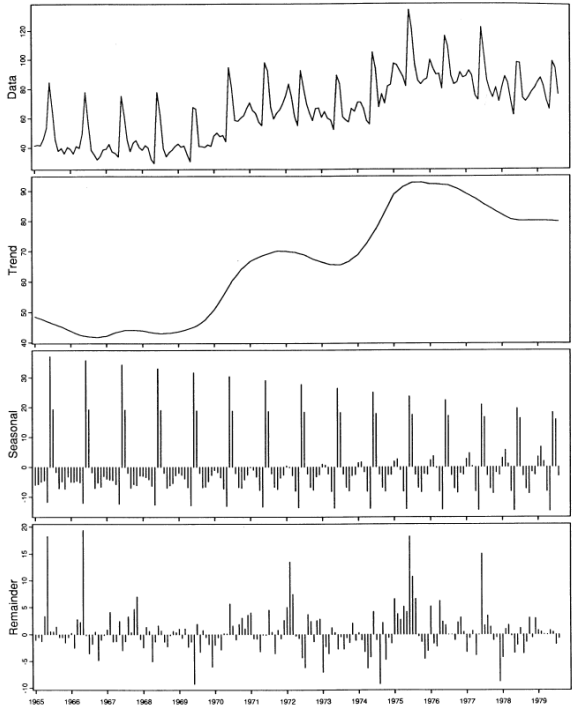

Fig. 4. An example of decomposing U.S unemployed males at ages 16-19 (Y-axis unit is 10,000)

From the algorithm perspective, STL is done using Loess smoothing. Loess smoothing is a convolutional smoothing filter taking different time points around *t* into consideration by applying different weighting factors. Time points far away from *t* will be assigned with smaller weighting factors and vice versa. An inner loop, outer loop and a low-pass filter are used in conjunction with Loess smoothing to complete the decomposition process. An example of STL to decompose U.S unemployment data is shown in Fig.4 [8].

### B. Support Vector Machine

Support Vector Machine (SVM) has been used as a powerful binary classifier in many applications [9]. It is chosen in this method. Fundamentally, A SVM automatically constructs a hyperplane to divide data points to two groups with maximum margin (i.e. the optimal hyperplane). In the original dimensional space, the data points are often not linearly separable. By using specific kernel functions, SVM project data points into higher dimensional space, presuming the higher dimensional space can provide better linear separablility. There are four basic kernel functions often used by SVM:

- Linear kernel: $k(\vec{X_1}, \vec{X_2}) = \vec{X_1} \cdot \vec{X_2}$;
- Polynomial kernel: $k(\vec{X_1}, \vec{X_2}) = (\vec{X_1} \cdot \vec{X_2})^d$, $d$ is the polynomial degree;
- Gaussian radial basis function kernel:

$$k(\vec{X_1}, \vec{X_2}) = exp\left(-\gamma \|\vec{X_1} - \vec{X_2}\|^2\right), \gamma > 0 \quad (2)$$

- Sigmoid function kernel:

$$k(\vec{X_1}, \vec{X_2}) = tanh(\vec{X_1} \cdot \vec{X_2}), \gamma > 0 \quad (3)$$

In Section IV, each of the above SVM kernels is put into test for classifying features produced from the residual components generated by STL modules.

## IV. THE APPLICATION RESULTS

The proposed method shown in Fig.2 was applied to all 8,711 signals included in the VSB dataset. STL with four different lengths of seasonal windows was applied to each waveform signal to produce four different STL decompositions. The lengths of the window were set to be 100, 1000, 10000 and 50000 points. An example of the decomposition result is shown in Fig.5.

To characterize the residual component, three features are proposed: the sum of absolute values, the maximum of absolute values and the standard deviation of absolute values. Absolute values are used because PD is such a random event that there is no actual pattern difference between the positive and negative values in the residual component. Fig.6 shows the 3D plot of the three features using 100-point seasonal window length. As can been seen, in general, the PD signals (red) are above the non-PD signals (green). The purpose of SVM is find the optimal hyperplane to separate these points. Since four STL modules are used, in total 12 features are produced for each signal. In the end, the 12 features are combined together and normalized to the range of [0,1] by Max-Min scaling to avoid feature biases [10].

One necessary technique that is required during the training of SVM is oversampling of PD records. This is because the numbers of PD signals are much fewer than non-PD signals in the dataset. Directly training could result in biased classification towards non-PD signals. One technique that can overcome this problem is purposely duplicating the PD signals so that the numbers of PD signals and non-PD signals are approximately the same in the dataset.

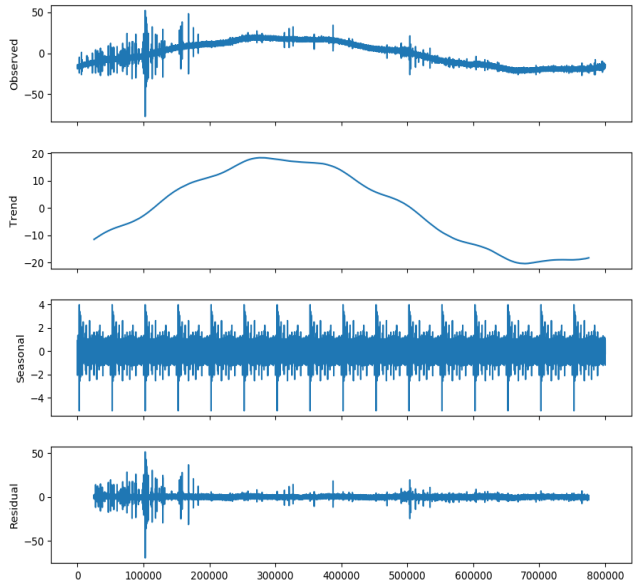

Fig. 5. An example of STL decomposition on a signal with PD

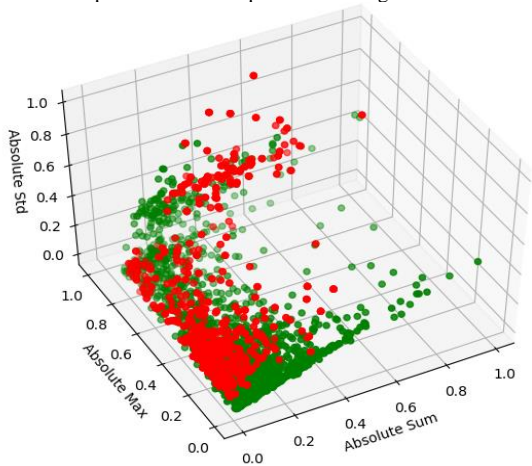

Fig. 6. 3D-Plot of PD signals (red) and non-PD signals (green)

Classification indexes Precision, Recall and F1-Score are used in the evaluation of different SVM kernels [11]. Their results are summarized in the tables below. As can be seen, the SVM with Gaussian radial basis kernel provided the best classification results by F1-Score in all categories. It is therefore selected as the final classifier for future application.

TABLE I. PRECISION, RECALL AND F1-SCORE FOR PREDICTION EVALUATION (LINEAR SVM KERNAL)

| Evaluation Category | Precision | Recall | F1-Score |
|---|---|---|---|
| Non-PD Signals | 0.77 | 0.58 | 0.66 |
| PD Signals | 0.67 | 0.83 | 0.74 |
| Average | 0.72 | 0.70 | 0.70 |

TABLE II. PRECISION, RECALL AND F1-SCORE FOR PREDICTION EVALUATION (POLYNOMIAL SVM KERNAL WITH DEGREE 6)

| Evaluation Category | Precision | Recall | F1-Score |
|---|---|---|---|
| Non-PD Signals | 0.52 | 0.97 | 0.68 |
| PD Signals | 0.82 | 0.13 | 0.23 |
| Average | 0.67 | 0.55 | 0.45 |

TABLE III. PRECISION, RECALL AND F1-SCORE FOR PREDICTION EVALUATION (GAUSSIAN RADIAL BASIS FUNCTION KERNEL)

| Evaluation Category | Precision | Recall | F1-Score |
|---|---|---|---|
| Non-PD Signals | 0.82 | 0.57 | 0.68 |
| PD Signals | 0.68 | 0.88 | 0.77 |
| Average | 0.75 | 0.73 | 0.72 |

TABLE IV. PRECISION, RECALL AND F1-SCORE FOR PREDICTION EVALUATION (SIGMOID FUNCTION KERNEL)

| Evaluation Category | Precision | Recall | F1-Score |
|---|---|---|---|
| Non-PD Signals | 0.51 | 0.74 | 0.60 |
| PD Signals | 0.54 | 0.30 | 0.39 |
| Average | 0.53 | 0.52 | 0.50 |

As Table III shows, 88% of actual PD signals can be successfully detected and 68% of detected PD signals are true PD signals.

## V. CONCLUSIONS AND FUTURE WORK

This paper developed a novel pattern recognition method to detect partial discharge activities on insulated overhead conductors based on stray electrical field voltage measurement. This method uses multiple STL modules to engineer multiple residual component features. The method was tested on the VSB dataset and reasonable detection performance was achieved.

Future work should include evaluation of other classification models such as neural network and naive Bayes classifier [10]. Also other noise extraction algorithms such as wavelet transformation can be compared with STL.